\documentstyle{article}

\makeatletter

\def\@aabuffer{}
\def\author #1{\expandafter\def\expandafter\@aabuffer\expandafter
{\@aabuffer \small\rm      #1\relax \par}}
\def\address#1{\expandafter\def\expandafter\@aabuffer\expandafter
{\@aabuffer \small\it #1\relax \par\vspace{1em}}}

\def\maketitle{
\begin{center}
   {\bf \@title \par}       
   \vskip 2em                      
   \@aabuffer\relax
\end{center} \par
\gdef\@aabuffer{}
}

\def\abstracts#1{
\begin{center}
{\begin{minipage}{4.2truein}
                 \footnotesize
                 \parindent=0pt #1\par
                 \end{minipage}}\end{center}
                 \vskip 2em \par}

\fussy
\def\section{\@startsection {section}{1}{\z@}{-3.25ex plus -1ex minus 
    -.2ex}{1.5ex plus .2ex}{\bf }}
\def\subsection{\@startsection{subsection}{2}{\z@}{-3.25ex plus -1ex minus 
    -.2ex}{1.5ex plus .2ex}{\it }}

\def\@makefnmark{{$\!^{\@thefnmark}$}}

\renewenvironment{thebibliography}[1]
	{\begin{list}{\arabic{enumi}.}
	{\usecounter{enumi}\setlength{\parsep}{0pt}
	 \setlength{\itemsep}{0pt} 
         \settowidth
	{\labelwidth}{#1.}\sloppy}}{\end{list}}

\newcounter{arabiclistc}

\def\@citex[#1]#2{\if@filesw\immediate\write\@auxout 
	{\string\citation{#2}}\fi 
\def\@citea{}\@cite{\@for\@citeb:=#2\do 
	{\@citea\def\@citea{,}\@ifundefined 
	{b@\@citeb}{{\bf ?}\@warning 
	{Citation `\@citeb' on page \thepage \space undefined}} 
	{\csname b@\@citeb\endcsname}}}{#1}} 
 
\newif\if@cghi 
\def\cite{\@cghitrue\@ifnextchar [{\@tempswatrue 
	\@citex}{\@tempswafalse\@citex[]}} 
\def\citelow{\@cghifalse\@ifnextchar [{\@tempswatrue 
	\@citex}{\@tempswafalse\@citex[]}} 
\def\@cite#1#2{{$\!^{#1}$\if@tempswa\typeout 
	{IJCGA warning: optional citation argument  
	ignored: `#2'} \fi}}

\def\baselinestretch{1.0} 
\ifx\selectfont\undefined 
\@normalsize\else\let\glb@currsize=\relax\selectfont 
\fi 
 
\ifx\selectfont\undefined 
\def\@singlespacing{%
\def\baselinestretch{1}\ifx\@currsize\normalsize\@normalsize\else\@currsize\fi%
} 
\else 
\def\@singlespacing{\def\baselinestretch{1}\let\glb@currsize=\relax\selectfont} 
\fi 
 
\long\def\@makecaption#1#2{ 
   \vskip 10pt  
   \setbox\@tempboxa\hbox{\footnotesize #1: #2} 
 \ifdim \wd\@tempboxa >\hsize \footnotesize #1: #2\par \else \hbox 
 to\hsize{\hfil\box\@tempboxa\hfil}   
   \fi} 
 
\def\fileversion{v1.20a}
\def\filedate{21.6.94}
\edef\epsfigRestoreAt{\catcode`@=\number\catcode`@\relax}%
\catcode`\@=11\relax
\ifx\undefined\@makeother                
\def\@makeother#1{\catcode`#1=12\relax}  
\fi                                      
\immediate\write16{Document style option `epsfig', \fileversion\space
<\filedate> (edited by SPQR + pks)}
\newcount\EPS@Height \newcount\EPS@Width \newcount\EPS@xscale
\newcount\EPS@yscale
\def\psfigdriver#1{%
  \bgroup\edef\next{\def\noexpand\tempa{#1}}%
    \uppercase\expandafter{\next}%
    \def\LN{DVITOLN03}%
    \def\DVItoPS{DVITOPS}%
    \def\DVIPS{DVIPS}%
    \def\emTeX{EMTEX}%
    \def\OzTeX{OZTEX}%
    \def\Textures{TEXTURES}%
    \global\chardef\fig@driver=0
    \ifx\tempa\LN
        \global\chardef\fig@driver=0\fi
    \ifx\tempa\DVItoPS
        \global\chardef\fig@driver=1\fi
    \ifx\tempa\DVIPS
        \global\chardef\fig@driver=2\fi
    \ifx\tempa\emTeX
        \global\chardef\fig@driver=3\fi
    \ifx\tempa\OzTeX
        \global\chardef\fig@driver=4\fi
    \ifx\tempa\Textures
        \global\chardef\fig@driver=5\fi
  \egroup
\def\psfig@start{}%
\def\psfig@end{}%
\def\epsfig@gofer{}%
\ifcase\fig@driver
\typeout{WARNING! ****
 no specials for LN03 psfig}%
\or 
\def\psfig@start{}%
\def\psfig@end{\special{dvitops: import \@p@sfilefinal \space
\@p@swidth sp \space \@p@sheight sp \space fill}%
\if@clip \typeout{Clipping not supported}\fi
\if@angle \typeout{Rotating not supported}\fi
}%
\let\epsfig@gofer\psfig@end
\or 
\def\psfig@start{\special{ps::[begin]  \@p@swidth \space \@p@sheight \space%
        \@p@sbbllx \space \@p@sbblly \space%
        \@p@sbburx \space \@p@sbbury \space%
        startTexFig \space }%
        \if@clip
                \if@verbose
                        \typeout{(clipped to BB) }%
                \fi
                \special{ps:: doclip \space }%
        \fi
        \if@angle              
                \special {ps:: \@p@sangle \space rotate \space}
        \fi
        \special{ps: plotfile \@p@sfilefinal \space }%
        \special{ps::[end] endTexFig \space }%
}%
\def\psfig@end{}%
\def\epsfig@gofer{\if@clip
                        \if@verbose
                           \typeout{(clipped to BB)}%
                        \fi
                        \epsfclipon
                  \fi
                  \epsfsetgraph{\@p@sfilefinal}%
}%
\or 
\typeout{WARNING. You must have a .bb info file with the Bounding Box
  of the pcx file}%
\def\psfig@start{}%
\def\psfig@end{\typeout{pcx import of \@p@sfilefinal}%
\if@clip \typeout{Clipping not supported}\fi
\if@angle \typeout{Rotating not supported}\fi
\raisebox{\@p@srheight sp}{\special{em: graph \@p@sfilefinal}}}%
\def\epsfig@gofer{}%
\or 
\def\psfig@start{}%
\def\psfig@end{%
\EPS@Width\@p@swidth
\EPS@Height\@p@sheight
\divide\EPS@Width by 65781  
\divide\EPS@Height by 65781
\special{epsf=\@p@sfilefinal
\space
width=\the\EPS@Width
\space
height=\the\EPS@Height
}%
\if@clip \typeout{Clipping not supported}\fi
\if@angle \typeout{Rotating not supported}\fi
}%
\let\epsfig@gofer\psfig@end
\or 
\def\psfig@end{
         \EPS@Width=\@bbw  
         \divide\EPS@Width by 1000
         \EPS@xscale=\@p@swidth \divide \EPS@xscale by \EPS@Width
         \EPS@Height=\@bbh  
         \divide\EPS@Height by 1000
         \EPS@yscale=\@p@sheight \divide \EPS@yscale by\EPS@Height
  \ifnum\EPS@xscale>\EPS@yscale\EPS@xscale=\EPS@yscale\fi
\if@clip
   \if@verbose
      \typeout{(clipped to BB)}%
   \fi
   \epsfclipon
\fi
\special{illustration \@p@sfilefinal\space scaled \the\EPS@xscale}%
}%
\def\psfig@start{}%
\let\epsfig\psfig
\else
\typeout{WARNING. *** unknown  driver - no psfig}%
\fi
}%
\newdimen\ps@dimcent
\ifx\undefined\fbox
\newdimen\fboxrule
\newdimen\fboxsep
\newdimen\ps@tempdima
\newbox\ps@tempboxa
\long\def\fbox#1{\leavevmode\setbox\ps@tempboxa\hbox{#1}\ps@tempdima\fboxrule
    \advance\ps@tempdima \fboxsep \advance\ps@tempdima \dp\ps@tempboxa
   \hbox{\lower \ps@tempdima\hbox
  {\vbox{\hrule height \fboxrule
          \hbox{\vrule width \fboxrule \hskip\fboxsep
          \vbox{\vskip\fboxsep \box\ps@tempboxa\vskip\fboxsep}\hskip
                 \fboxsep\vrule width \fboxrule}%
                 \hrule height \fboxrule}}}}%
\fi
\ifx\@ifundefined\undefined
\long\def\@ifundefined#1#2#3{\expandafter\ifx\csname
  #1\endcsname\relax#2\else#3\fi}%
\fi
\@ifundefined{typeout}%
{\gdef\typeout#1{\immediate\write\sixt@@n{#1}}}%
{\relax}%
\@ifundefined{epsfig}{}{\typeout{EPSFIG --- already loaded}\endinput}%
\@ifundefined{epsfbox}{\input epsf}{}%
\ifx\undefined\@latexerr
        \newlinechar`\^^J
        \def\@spaces{\space\space\space\space}%
        \def\@latexerr#1#2{%
        \edef\@tempc{#2}\expandafter\errhelp\expandafter{\@tempc}%
        \typeout{Error. \space see a manual for explanation.^^J
         \space\@spaces\@spaces\@spaces Type \space H <return> \space for
         immediate help.}\errmessage{#1}}%
\fi
\def\@whattodo{You tried to include a PostScript figure which
cannot be found^^JIf you press return to carry on anyway,^^J
The failed name will be printed in place of the figure.^^J
or type X to quit}%
\def\@whattodobb{You tried to include a PostScript figure which
has no^^Jbounding box, and you supplied none.^^J
If you press return to carry on anyway,^^J
The failed name will be printed in place of the figure.^^J
or type X to quit}%
\def\@nnil{\@nil}%
\def\@empty{}%
\def\@psdonoop#1\@@#2#3{}%
\def\@psdo#1:=#2\do#3{\edef\@psdotmp{#2}\ifx\@psdotmp\@empty \else
    \expandafter\@psdoloop#2,\@nil,\@nil\@@#1{#3}\fi}%
\def\@psdoloop#1,#2,#3\@@#4#5{\def#4{#1}\ifx #4\@nnil \else
       #5\def#4{#2}\ifx #4\@nnil \else#5\@ipsdoloop #3\@@#4{#5}\fi\fi}%
\def\@ipsdoloop#1,#2\@@#3#4{\def#3{#1}\ifx #3\@nnil
       \let\@nextwhile=\@psdonoop \else
      #4\relax\let\@nextwhile=\@ipsdoloop\fi\@nextwhile#2\@@#3{#4}}%
\def\@tpsdo#1:=#2\do#3{\xdef\@psdotmp{#2}\ifx\@psdotmp\@empty \else
    \@tpsdoloop#2\@nil\@nil\@@#1{#3}\fi}%
\def\@tpsdoloop#1#2\@@#3#4{\def#3{#1}\ifx #3\@nnil
       \let\@nextwhile=\@psdonoop \else
      #4\relax\let\@nextwhile=\@tpsdoloop\fi\@nextwhile#2\@@#3{#4}}%
\long\def\epsfaux#1#2:#3\\{\ifx#1\epsfpercent
   \def\testit{#2}\ifx\testit\epsfbblit
        \@atendfalse
        \epsf@atend #3 . \\%
        \if@atend
           \if@verbose
                \typeout{epsfig: found `(atend)'; continuing search}%
           \fi
        \else
                \epsfgrab #3 . . . \\%
                \epsffileokfalse\global\no@bbfalse
                \global\epsfbbfoundtrue
        \fi
   \fi\fi}%
\def\epsf@atendlit{(atend)}
\def\epsf@atend #1 #2 #3\\{%
   \def\epsf@tmp{#1}\ifx\epsf@tmp\empty
      \epsf@atend #2 #3 .\\\else
   \ifx\epsf@tmp\epsf@atendlit\@atendtrue\fi\fi}%

\chardef\trig@letter = 11
\chardef\other = 12
 
\newif\ifdebug 
\newif\ifc@mpute 
\newif\if@atend
\c@mputetrue 
 
\let\then = \relax
\def\r@dian{pt }%
\let\r@dians = \r@dian
\let\dimensionless@nit = \r@dian
\let\dimensionless@nits = \dimensionless@nit
\def\internal@nit{sp }%
\let\internal@nits = \internal@nit
\newif\ifstillc@nverging
\def \Mess@ge #1{\ifdebug \then \message {#1} \fi}%
 
{ 
        \catcode `\@ = \trig@letter
        \gdef \nodimen {\expandafter \n@dimen \the \dimen}%
        \gdef \term #1 #2 #3%
               {\edef \t@ {\the #1}
                \edef \t@@ {\expandafter \n@dimen \the #2\r@dian}%
                \t@rm {\t@} {\t@@} {#3}%
               }%
        \gdef \t@rm #1 #2 #3%
               {{%
                \count 0 = 0
                \dimen 0 = 1 \dimensionless@nit
                \dimen 2 = #2\relax
                \Mess@ge {Calculating term #1 of \nodimen 2}%
                \loop
                \ifnum  \count 0 < #1
                \then   \advance \count 0 by 1
                        \Mess@ge {Iteration \the \count 0 \space}%
                        \Multiply \dimen 0 by {\dimen 2}%
                        \Mess@ge {After multiplication, term = \nodimen 0}%
                        \Divide \dimen 0 by {\count 0}%
                        \Mess@ge {After division, term = \nodimen 0}%
                \repeat
                \Mess@ge {Final value for term #1 of
                                \nodimen 2 \space is \nodimen 0}%
                \xdef \Term {#3 = \nodimen 0 \r@dians}%
                \aftergroup \Term
               }}%
        \catcode `\p = \other
        \catcode `\t = \other
        \gdef \n@dimen #1pt{#1} 
}%
 
\def \Divide #1by #2{\divide #1 by #2} 
 
\def \Multiply #1by #2
       {{
        \count 0 = #1\relax
        \count 2 = #2\relax
        \count 4 = 65536
        \Mess@ge {Before scaling, count 0 = \the \count 0 \space and
                        count 2 = \the \count 2}%
        \ifnum  \count 0 > 32767 
        \then   \divide \count 0 by 4
                \divide \count 4 by 4
        \else   \ifnum  \count 0 < -32767
                \then   \divide \count 0 by 4
                        \divide \count 4 by 4
                \else
                \fi
        \fi
        \ifnum  \count 2 > 32767 
        \then   \divide \count 2 by 4
                \divide \count 4 by 4
        \else   \ifnum  \count 2 < -32767
                \then   \divide \count 2 by 4
                        \divide \count 4 by 4
                \else
                \fi
        \fi
        \multiply \count 0 by \count 2
        \divide \count 0 by \count 4
        \xdef \product {#1 = \the \count 0 \internal@nits}%
        \aftergroup \product
       }}%
 
\def\r@duce{\ifdim\dimen0 > 90\r@dian \then   
                \multiply\dimen0 by -1
                \advance\dimen0 by 180\r@dian
                \r@duce
            \else \ifdim\dimen0 < -90\r@dian \then  
                \advance\dimen0 by 360\r@dian
                \r@duce
                \fi
            \fi}%
 
\def\Sine#1%
       {{%
        \dimen 0 = #1 \r@dian
        \r@duce
        \ifdim\dimen0 = -90\r@dian \then
           \dimen4 = -1\r@dian
           \c@mputefalse
        \fi
        \ifdim\dimen0 = 90\r@dian \then
           \dimen4 = 1\r@dian
           \c@mputefalse
        \fi
        \ifdim\dimen0 = 0\r@dian \then
           \dimen4 = 0\r@dian
           \c@mputefalse
        \fi
        \ifc@mpute \then
                \divide\dimen0 by 180
                \dimen0=3.141592654\dimen0
                \dimen 2 = 3.1415926535897963\r@dian 
                \divide\dimen 2 by 2 
                \Mess@ge {Sin: calculating Sin of \nodimen 0}%
                \count 0 = 1 
                \dimen 2 = 1 \r@dian 
                \dimen 4 = 0 \r@dian 
                \loop
                        \ifnum  \dimen 2 = 0 
                        \then   \stillc@nvergingfalse
                        \else   \stillc@nvergingtrue
                        \fi
                        \ifstillc@nverging 
                        \then   \term {\count 0} {\dimen 0} {\dimen 2}%
                                \advance \count 0 by 2
                                \count 2 = \count 0
                                \divide \count 2 by 2
                                \ifodd  \count 2 
                                \then   \advance \dimen 4 by \dimen 2
                                \else   \advance \dimen 4 by -\dimen 2
                                \fi
                \repeat
        \fi
                        \xdef \sine {\nodimen 4}%
       }}%
 
\def\Cosine#1{\ifx\sine\UnDefined\edef\Savesine{\relax}\else
                             \edef\Savesine{\sine}\fi
        {\dimen0=#1\r@dian\multiply\dimen0 by -1
         \advance\dimen0 by 90\r@dian
         \Sine{\nodimen 0}%
         \xdef\cosine{\sine}%
         \xdef\sine{\Savesine}}}
\def\psdraft{\def\@psdraft{0}}%
\def\psfull{\def\@psdraft{1}}%
\psfull
\newif\if@compress
\def\pscompress{\@compresstrue}
\def\psnocompress{\@compressfalse}
\@compressfalse
\newif\if@scalefirst
\def\psscalefirst{\@scalefirsttrue}%
\def\psrotatefirst{\@scalefirstfalse}%
\psrotatefirst
\newif\if@draftbox
\def\psnodraftbox{\@draftboxfalse}%
\@draftboxtrue
\newif\if@noisy
\@noisyfalse
\newif\ifno@bb
\newif\if@bbllx
\newif\if@bblly
\newif\if@bburx
\newif\if@bbury
\newif\if@height
\newif\if@width
\newif\if@rheight
\newif\if@rwidth
\newif\if@angle
\newif\if@clip
\newif\if@verbose
\newif\if@prologfile
\def\@p@@sprolog#1{\@prologfiletrue\def\@prologfileval{#1}}%
\def\@p@@sclip#1{\@cliptrue}%
\newif\ifepsfig@dos  
\def\epsfigdos{\epsfig@dostrue}%
\epsfig@dosfalse
\newif\ifuse@psfig
\def\ParseName#1{\expandafter\@Parse#1}%
\def\@Parse#1.#2:{\gdef\BaseName{#1}\gdef\FileType{#2}}%

\def\@p@@sfile#1{%
  \ifepsfig@dos
     \ParseName{#1:}%
  \else
     \gdef\BaseName{#1}\gdef\FileType{}%
  \fi
  \def\@p@sfile{NO FILE: #1}%
  \def\@p@sfilefinal{NO FILE: #1}%
  \openin1=#1
  \ifeof1\closein1\openin1=\BaseName.bb
    \ifeof1\closein1
      \if@bbllx                 
        \if@bblly\if@bburx\if@bbury
          \def\@p@sfile{#1}%
          \def\@p@sfilefinal{#1}%
        \fi\fi\fi
      \else                     
        \@latexerr{ERROR. PostScript file #1 not found}\@whattodo
        \@p@@sbbllx{100bp}%
        \@p@@sbblly{100bp}%
        \@p@@sbburx{200bp}%
        \@p@@sbbury{200bp}%
        \psdraft
      \fi
    \else                       
      \closein1%
      \edef\@p@sfile{\BaseName.bb}%
      \typeout{using BB from \@p@sfile}%
      \ifnum\fig@driver=3
        \edef\@p@sfilefinal{\BaseName.pcx}%
      \else
        \ifepsfig@dos
          \edef\@p@sfilefinal{"`gunzip -c `texfind \BaseName.{z,Z,gz}"}%
        \else
          \edef\@p@sfilefinal{"`epsfig \if@compress-c \fi#1"}%
        \fi
      \fi
    \fi
  \else\closein1                
    \edef\@p@sfile{#1}%
    \if@compress  
      \edef\@p@sfilefinal{"`epsfig -c #1"}%
    \else
      \edef\@p@sfilefinal{#1}%
    \fi
  \fi%
}

\let\@p@@sfigure\@p@@sfile
\def\@p@@sbbllx#1{%
                                            \@bbllxtrue
                \ps@dimcent=#1
                \edef\@p@sbbllx{\number\ps@dimcent}%
                \divide\ps@dimcent by65536
                \global\edef\epsfllx{\number\ps@dimcent}%
}%
\def\@p@@sbblly#1{%
                \@bbllytrue
                \ps@dimcent=#1
                \edef\@p@sbblly{\number\ps@dimcent}%
                \divide\ps@dimcent by65536
                \global\edef\epsflly{\number\ps@dimcent}%
}%
\def\@p@@sbburx#1{%
                \@bburxtrue
                \ps@dimcent=#1
                \edef\@p@sbburx{\number\ps@dimcent}%
                \divide\ps@dimcent by65536
                \global\edef\epsfurx{\number\ps@dimcent}%
}%
\def\@p@@sbbury#1{%
                \@bburytrue
                \ps@dimcent=#1
                \edef\@p@sbbury{\number\ps@dimcent}%
                \divide\ps@dimcent by65536
                \global\edef\epsfury{\number\ps@dimcent}%
}%
\def\@p@@sheight#1{%
                \@heighttrue
                \global\epsfysize=#1
                \ps@dimcent=#1
                \edef\@p@sheight{\number\ps@dimcent}%
}%
\def\@p@@swidth#1{%
                \@widthtrue
                \global\epsfxsize=#1
                \ps@dimcent=#1
                \edef\@p@swidth{\number\ps@dimcent}%
}%
\def\@p@@srheight#1{%
                \@rheighttrue\use@psfigtrue
                \ps@dimcent=#1
                \edef\@p@srheight{\number\ps@dimcent}%
}%
\def\@p@@srwidth#1{%
                \@rwidthtrue\use@psfigtrue
                \ps@dimcent=#1
                \edef\@p@srwidth{\number\ps@dimcent}%
}%
\def\@p@@sangle#1{%
                \use@psfigtrue
                \@angletrue
                \edef\@p@sangle{#1}%
}%
\def\@p@@ssilent#1{%
                \@verbosefalse
}%
\def\@p@@snoisy#1{%
                \@verbosetrue
}%
\def\@cs@name#1{\csname #1\endcsname}%
\def\@setparms#1=#2,{\@cs@name{@p@@s#1}{#2}}%
\def\ps@init@parms{%
                \@bbllxfalse \@bbllyfalse
                \@bburxfalse \@bburyfalse
                \@heightfalse \@widthfalse
                \@rheightfalse \@rwidthfalse
                \def\@p@sbbllx{}\def\@p@sbblly{}%
                \def\@p@sbburx{}\def\@p@sbbury{}%
                \def\@p@sheight{}\def\@p@swidth{}%
                \def\@p@srheight{}\def\@p@srwidth{}%
                \def\@p@sangle{0}%
                \def\@p@sfile{}%
                \use@psfigfalse
                \@prologfilefalse
                \def\@sc{}%
                \if@noisy
                        \@verbosetrue
                \else
                        \@verbosefalse
                \fi
                \@clipfalse
}%
\def\parse@ps@parms#1{%
                \@psdo\@psfiga:=#1\do
                   {\expandafter\@setparms\@psfiga,}%
\if@prologfile
\fi
}%
\def\bb@missing{%
        \if@verbose
            \typeout{psfig: searching \@p@sfile \space  for bounding box}%
        \fi
        \epsfgetbb{\@p@sfile}%
        \ifepsfbbfound
            \ps@dimcent=\epsfllx bp\edef\@p@sbbllx{\number\ps@dimcent}%
            \ps@dimcent=\epsflly bp\edef\@p@sbblly{\number\ps@dimcent}%
            \ps@dimcent=\epsfurx bp\edef\@p@sbburx{\number\ps@dimcent}%
            \ps@dimcent=\epsfury bp\edef\@p@sbbury{\number\ps@dimcent}%
        \else
            \epsfbbfoundfalse
        \fi
}
\newdimen\p@intvaluex
\newdimen\p@intvaluey
\def\rotate@#1#2{{\dimen0=#1 sp\dimen1=#2 sp
                  \global\p@intvaluex=\cosine\dimen0
                  \dimen3=\sine\dimen1
                  \global\advance\p@intvaluex by -\dimen3
                  \global\p@intvaluey=\sine\dimen0
                  \dimen3=\cosine\dimen1
                  \global\advance\p@intvaluey by \dimen3
                  }}%
\def\compute@bb{%
                \epsfbbfoundfalse
                \if@bbllx\epsfbbfoundtrue\fi
                \if@bblly\epsfbbfoundtrue\fi
                \if@bburx\epsfbbfoundtrue\fi
                \if@bbury\epsfbbfoundtrue\fi
                \ifepsfbbfound\else\bb@missing\fi
                \ifepsfbbfound\else
                \@latexerr{ERROR. cannot locate BoundingBox}\@whattodobb
                        \@p@@sbbllx{100bp}%
                        \@p@@sbblly{100bp}%
                        \@p@@sbburx{200bp}%
                        \@p@@sbbury{200bp}%
                        \no@bbtrue
                        \psdraft
                \fi
                \count203=\@p@sbburx
                \count204=\@p@sbbury
                \advance\count203 by -\@p@sbbllx
                \advance\count204 by -\@p@sbblly
                \edef\ps@bbw{\number\count203}%
                \edef\ps@bbh{\number\count204}%
                 \edef\@bbw{\number\count203}%
                \edef\@bbh{\number\count204}%
               \if@angle
                        \Sine{\@p@sangle}\Cosine{\@p@sangle}%
 
{\ps@dimcent=\maxdimen\xdef\r@p@sbbllx{\number\ps@dimcent}%
 
\xdef\r@p@sbblly{\number\ps@dimcent}%
 
\xdef\r@p@sbburx{-\number\ps@dimcent}%
 
\xdef\r@p@sbbury{-\number\ps@dimcent}}%
                        \def\minmaxtest{%
                           \ifnum\number\p@intvaluex<\r@p@sbbllx
                              \xdef\r@p@sbbllx{\number\p@intvaluex}\fi
                           \ifnum\number\p@intvaluex>\r@p@sbburx
                              \xdef\r@p@sbburx{\number\p@intvaluex}\fi
                           \ifnum\number\p@intvaluey<\r@p@sbblly
                              \xdef\r@p@sbblly{\number\p@intvaluey}\fi
                           \ifnum\number\p@intvaluey>\r@p@sbbury
                              \xdef\r@p@sbbury{\number\p@intvaluey}\fi
                           }%
                        \rotate@{\@p@sbbllx}{\@p@sbblly}%
                        \minmaxtest
                        \rotate@{\@p@sbbllx}{\@p@sbbury}%
                        \minmaxtest
                        \rotate@{\@p@sbburx}{\@p@sbblly}%
                        \minmaxtest
                        \rotate@{\@p@sbburx}{\@p@sbbury}%
                        \minmaxtest
 
\edef\@p@sbbllx{\r@p@sbbllx}\edef\@p@sbblly{\r@p@sbblly}%
 
\edef\@p@sbburx{\r@p@sbburx}\edef\@p@sbbury{\r@p@sbbury}%
                \fi
                \count203=\@p@sbburx
                \count204=\@p@sbbury
                \advance\count203 by -\@p@sbbllx
                \advance\count204 by -\@p@sbblly
                \edef\@bbw{\number\count203}%
                \edef\@bbh{\number\count204}%
}%
\def\in@hundreds#1#2#3{\count240=#2 \count241=#3
                     \count100=\count240        
                     \divide\count100 by \count241
                     \count101=\count100
                     \multiply\count101 by \count241
                     \advance\count240 by -\count101
                     \multiply\count240 by 10
                     \count101=\count240        
                     \divide\count101 by \count241
                     \count102=\count101
                     \multiply\count102 by \count241
                     \advance\count240 by -\count102
                     \multiply\count240 by 10
                     \count102=\count240        
                     \divide\count102 by \count241
                     \count200=#1\count205=0
                     \count201=\count200
                        \multiply\count201 by \count100
                        \advance\count205 by \count201
                     \count201=\count200
                        \divide\count201 by 10
                        \multiply\count201 by \count101
                        \advance\count205 by \count201
                     \count201=\count200
                        \divide\count201 by 100
                        \multiply\count201 by \count102
                        \advance\count205 by \count201
                     \edef\@result{\number\count205}%
}%
\def\compute@wfromh{%
                \in@hundreds{\@p@sheight}{\@bbw}{\@bbh}%
                \edef\@p@swidth{\@result}%
}%
\def\compute@hfromw{%
                \in@hundreds{\@p@swidth}{\@bbh}{\@bbw}%
                \edef\@p@sheight{\@result}%
}%
\def\compute@handw{%
                \if@height
                        \if@width
                        \else
                                \compute@wfromh
                        \fi
                \else
                        \if@width
                                \compute@hfromw
                        \else
                                \edef\@p@sheight{\@bbh}%
                                \edef\@p@swidth{\@bbw}%
                        \fi
                \fi
}%
\def\compute@resv{%
                \if@rheight \else \edef\@p@srheight{\@p@sheight} \fi
                \if@rwidth \else \edef\@p@srwidth{\@p@swidth} \fi
}%
\def\compute@sizes{%
        \if@scalefirst\if@angle
        \if@width
           \in@hundreds{\@p@swidth}{\@bbw}{\ps@bbw}%
           \edef\@p@swidth{\@result}%
        \fi
        \if@height
           \in@hundreds{\@p@sheight}{\@bbh}{\ps@bbh}%
           \edef\@p@sheight{\@result}%
        \fi
        \fi\fi
        \compute@handw
        \compute@resv
}
\long\def\graphic@verb#1{\def\next{#1}%
  {\expandafter\graphic@strip\meaning\next}}
\def\graphic@strip#1>{}
\def\graphic@zapspace#1{%
  #1\ifx\graphic@zapspace#1\graphic@zapspace%
  \else\expandafter\graphic@zapspace%
  \fi}
\def\psfig#1{%
\edef\@tempa{\graphic@zapspace#1{}}%
\ifvmode\leavevmode\fi\vbox {%
        \ps@init@parms
        \parse@ps@parms{\@tempa}%
        \ifnum\@psdraft=1
                \typeout{[\@p@sfilefinal]}%
                \if@verbose
                        \typeout{epsfig: using PSFIG macros}%
                \fi
                \psfig@method
        \else
                \epsfig@draft
        \fi
}
}%
\def\graphic@zapspace#1{%
  #1\ifx\graphic@zapspace#1\graphic@zapspace%
  \else\expandafter\graphic@zapspace%
  \fi}
\def\epsfig#1{%
\edef\@tempa{\graphic@zapspace#1{}}%
\ifvmode\leavevmode\fi\vbox {%
        \ps@init@parms
        \parse@ps@parms{\@tempa}%
        \ifnum\@psdraft=1
          \if@angle\use@psfigtrue\fi
          {\ifnum\fig@driver=1\global\use@psfigtrue\fi}%
          {\ifnum\fig@driver=3\global\use@psfigtrue\fi}%
          {\ifnum\fig@driver=4\global\use@psfigtrue\fi}%
          {\ifnum\fig@driver=5\global\use@psfigtrue\fi}%
                \ifuse@psfig
                        \if@verbose
                                \typeout{epsfig: using PSFIG macros}%
                        \fi
                        \psfig@method
                \else
                        \if@verbose
                                \typeout{epsfig: using EPSF macros}%
                        \fi
                        \epsf@method
                \fi
        \else
                \epsfig@draft
        \fi
}%
}%

\def\epsf@method{%
        \epsfbbfoundfalse
        \if@bbllx\epsfbbfoundtrue\fi
        \if@bblly\epsfbbfoundtrue\fi
        \if@bburx\epsfbbfoundtrue\fi
        \if@bbury\epsfbbfoundtrue\fi
        \ifepsfbbfound\else\epsfgetbb{\@p@sfile}\fi
        \ifepsfbbfound
           \typeout{<\@p@sfilefinal>}%
           \epsfig@gofer
        \else
          \@latexerr{ERROR - Cannot locate BoundingBox}\@whattodobb
          \@p@@sbbllx{100bp}%
          \@p@@sbblly{100bp}%
          \@p@@sbburx{200bp}%
          \@p@@sbbury{200bp}%
                \count203=\@p@sbburx
                \count204=\@p@sbbury
                \advance\count203 by -\@p@sbbllx
                \advance\count204 by -\@p@sbblly
                \edef\@bbw{\number\count203}%
                \edef\@bbh{\number\count204}%
          \compute@sizes
          \epsfig@@draft
       \fi
}%
\def\psfig@method{%
        \compute@bb
        \ifepsfbbfound
          \compute@sizes
          \psfig@start
          \vbox to \@p@srheight sp{\hbox to \@p@srwidth 
            sp{\hss}\vss\psfig@end}%
        \else
           \epsfig@draft
        \fi
}%
\def\epsfig@draft{\compute@bb\compute@sizes\epsfig@@draft}%
\def\epsfig@@draft{%
\typeout{<(draft only) \@p@sfilefinal>}%
\if@draftbox
        \hbox{{\fboxsep0pt\fbox{\vbox to \@p@srheight sp{%
        \vss\hbox to \@p@srwidth sp{ \hss 
           \expandafter\Literally\@p@sfilefinal\@nil
                          \hss }\vss
        }}}}%
\else
        \vbox to \@p@srheight sp{%
        \vss\hbox to \@p@srwidth sp{\hss}\vss}%
\fi
}%
\def\Literally#1\@nil{{\tt\graphic@verb{#1}}}
\psfigdriver{dvips}%
\epsfigRestoreAt

\makeatother

\def\Journal#1#2#3#4{{#1} {\bf #2}, #3 (#4)}

\def\PLB{{\em Phys. Lett.}  B}

\def\PRD{{\em Phys. Rev.} D}
\def\PRA{{\em Phys. Rev.} A}
\def\AP{{\em Ann. Phys.} (N.Y.)}

\def\be{\begin{equation}}
\def\ee{\end{equation}}
\def\bea{\begin{eqnarray}}
\def\eea{\end{eqnarray}}

\begin{document}

\sloppy

\setcounter{page}{0}
\pagestyle{empty}

\begin{flushright}
NORDITA--97/52 P/A  \\
hep-ph/9708383  \\
13 August 1997
\end{flushright}
\vspace*{32mm}

\title{EARLY STAGES OF GROWTH OF QCD AND ELECTROWEAK 
  BUBBLES\protect\footnotemark[1] \vspace*{8mm} }
\footnotetext[1]{
 To appear in the Proceedings of Strong and Electroweak Matter '97, Eger,
 Hungary, 21--25 May 1997.}

\author{ J. IGNATIUS \vspace*{3mm} }

\address{NORDITA, Blegdamsvej 17, DK-2100 Copenhagen \O, Denmark, and\\
  Department of Physics, FIN-00014 University of Helsinki, Finland. \\
  E-mail: ignatius@iki.fi \vspace*{12mm} }


\maketitle\abstracts{
The dynamical growth rate of bubbles nucleating in relativistic plasma in
thermal first-order phase transitions is analyzed. The framework is a
hydrodynamical model which consists of relativistic fluid and an order 
parameter field. The results of analytical approximations and numerical 
simulations coincide well. 
}

\clearpage

\pagestyle{plain}

In thermal systems first-order phase transitions normally proceed via 
nucleation  of bubbles of the new phase. Bubbles larger than a
certain critical size begin to grow, whereas smaller bubbles will shrink. 
Langer's formula~\cite{Langer69}$^{\!,\,}$\cite{Langer73} for the nucleation
rate of bubbles of the new phase is given by
\begin{equation}
  \Gamma = \frac{\kappa}{2\pi} \Omega_0 \: e^{-\Delta F/T}, \label{Gamma}
\end{equation}
where $\Gamma$ is the probability of nucleation per volume and time, $\kappa$
a dynamical and $\Omega_0$ a statistical prefactor, $\Delta F$ the free energy
difference of the system with and without the nucleating bubble, and $T$ the
temperature.  The purpose of this study is to investigate the dynamical
prefactor $\kappa$.

Let the phase transition be driven by an order parameter field
$\phi(x)$. Initial growth rate of perturbations around an extremum
configuration is given by the coefficient
$\kappa$:
\begin{equation}
  \delta \phi \; \propto \; e^{\kappa t}.  \label{deltaphi}
\end{equation}
Similar relations hold for perturbations of energy density and fluid velocity.
A parallel definition for the growth coefficient $\kappa$ can be expressed in
terms of the radius $R(t)$ of an expanding spherical bubble of the new phase,
\begin{equation}
  \frac{{\rm d}R(t)}{{\rm d}t} \approx \kappa [R(t) - R_{\rm cr}],
 \label{rkdef}
\end{equation}
where $R_{\rm cr}$ is the radius of the critical or extremum bubble. This
equation is valid when $R(t) \approx R_{\rm cr}$. In order for the bubble to
grow, the initial radius $R(0)$ must be slightly larger than that of a
critical bubble.

In literature there are at least two calculations of the growth rate $\kappa$
for relativistic plasma. Csernai and Kapusta write the free energy density as
a functional of their order parameter which is just the usual internal energy
density.\cite{Csernai92} Ruggeri and Friedman make the approximation that the
interface is infinitely thin and do not need any order
parameter.\cite{Ruggeri96} These two methods produce results
for $\kappa$ which disagree even qualitatively with each other.

Let us now introduce a hydrodynamical model which enables both an analytical
and a numerical determination of the initial growth rate $\kappa$. The
model~\cite{Ignatius94} consists of an order parameter field $\phi$, and
perfect fluid which describes the other degrees of freedom.  There are three
basic locally varying quantities, namely $\phi(x)$, fluid four-velocity
$u^{\mu}(x)$, and temperature $T(x)$. Due to the small value of the baryon
asymmetry all the conserved particle numbers can for the present purposes well
be approximated to be zero in the early Universe. Furthermore, the expansion
of the Universe can be neglected, since the whole period of nucleation, yet
alone the initial growth of bubbles, is an extremely rapid process.  Thus the
equations of motion can be written in the form
\begin{equation}
  \left\{ \begin{array}{l} \partial_{\mu} T^{\mu\nu} = 0 \\ \partial^2 \phi +
    \frac{\partial}{\partial\phi} V(\phi,T) = - \eta \, u^{\mu} \partial_{\mu}
    \phi, \\ \end{array} \right.  \label{EOM}
\end{equation}
where $T^{\mu\nu} = T^{\mu\nu}\left\{u^{\alpha}(x),
\partial_{\beta}\phi(x),\phi(x),T(x)\right\}$ is the energy-momentum tensor, 
$V$ the potential energy density for the order parameter field, $T$ the
temperature, and $\eta$ a phenomenological friction parameter (not to be
confused with shear viscosity).  The upper equation is the conservation law of
total energy-momentum. The lower equation tells how energy is transported
between the order parameter and the fluid through the dissipative term,
proportional to $\eta$. The same term is also responsible for the creation of
entropy. In the limit of vanishing fluid velocities the lower equation gives
the simple dissipative equation
\begin{equation}
  \frac{{\rm d}\phi}{{\rm d}t} = - \frac{1}{\eta} \frac{\delta
      S_3[\phi]}{\delta \phi}, \label{dphidt}
\end{equation}
where $S_3[\phi]$ is the usual high-temperature three-dimensional action
(equalling free energy at extrema).

It is clear that a hydrodynamical description of the system cannot be
complete. It is only valid at scales which are longer than particle mean free
paths or interaction times. The solutions to Eqs.~(\ref{EOM}) are smoothly
behaving fields, whereas in reality fields have strong thermal fluctuations on
short scales. To incorporate thermal fluctuations a Langevin-type equation
with a noise term would be needed. Within a purely hydrodynamical model
questions for example about damping effects in plasma cannot be answered.

The model in Eqs.~(\ref{EOM}) can be applied to describe both electroweak and
QCD phase transition. In electroweak theory the order parameter is obviously
identified as the Higgs field. The true coupling term is more complicated than
the frictional $\eta$-term of this model. But that effect should not be
significant, as long as the detailed internal structure of the interface is
not being discussed (something which a hydrodynamical model cannot accurately
determine anyway). The value of the friction parameter $\eta$ can be fixed by
comparing with microscopic calculations.\cite{Moore95} In the case of QCD the
order parameter cannot be identified with a physical particle.  However, one
can still employ the model as a purely phenomenological
construction. Interaction length or time of QCD sets the scale for the
friction coefficient $\eta$. A further complication is the macroscopic mean
free path of neutrinos, but luckily the hydrodynamic energy flux is in normal
cases clearly superior compared with that carried by the neutrinos.

For solving Eqs.~(\ref{EOM}) the potential $V(\phi,T)$ must be known. Here the
usual quartic potential has been employed. By fixing the parameters of it in a
suitable manner, the desired values for the surface tension and latent heat of
the transition will be reproduced. The coordinate system is spherically
symmetric 1+3 dimensional space-time.\cite{KurkiSuonio96}

In order to create the initial configuration the critical bubble solution must
be known with high accuracy. Going closer to the thin-wall limit, that is,
using larger critical bubbles, has the advantage that numerical errors in the
determination of growth coefficient $\kappa$ decrease. But in this limit the
field equation cannot be directly integrated numerically to produce the
critical bubble. Instead, the following Ansatz is used:
\begin{equation}
 \phi_A (r) = \frac{\phi_{\rm min }}{2} \left\{ 1 
    - \tanh \left( \frac{r-R_{\rm cr}}{2\xi_{\rm cr}} \right) \right\}.
\end{equation}
Here $R_{\rm cr}$, $\xi_{\rm cr}$ are {\em unknown} parameters, 
and $\phi_{\rm min}$ is the position of the new minimum of the potential. 
In this two-dimensional
subspace set by the Ansatz the extremum of the action becomes saddle point of 
an ordinary function, $S_3[\phi_A] = S_{3A} (R_{\rm cr}, \xi_{\rm cr})$. This
saddle point can then be located numerically. For larger bubbles this method
produces quite accurate results, which was actually unexpected. Fixing $R_{\rm
cr}$ naively by Laplace's relation, $R_{\rm cr} = 2 \sigma / \Delta p$, where
$\sigma$ is surface tension and $\Delta p$ pressure difference between the two
phases, would lead to huge inaccuracies. However, the correct value of $R_{\rm
cr}$ can be found directly by employing curvature-dependent surface
tension~\cite{Ignatius93}, $\sigma(R)$.
 
Analytically the initial growth rate $\kappa$ can be determined as follows.
Expand the low-velocity dissipation equation (\ref{dphidt}) around the
critical bubble $\bar{\phi}$ by making the substitution $ \phi(t,{\bf x}) =
\bar{\phi}(r) + \varphi(t,{\bf x})$. The resulting equation for fluctuations
is
\begin{equation}
 \frac{{\rm d}\varphi}{{\rm d}t} =
 - \frac{1}{\eta} \frac{ \delta^2 S_3[\bar{\phi}] }{\delta^2 \phi} \varphi.
\end{equation}
Next insert the unstable growth mode, Eq.~(\ref{deltaphi}). The thin-wall
result for the negative eigenvalue of the fluctuation operator, $\lambda_-
= 2/R_{\rm cr}^2$, has been observed to hold surprisingly well in the general
case, too.\cite{Brihaye93} This gives the approximate solution for the initial
growth rate in the hydrodynamical model:
\begin{equation}
  \kappa \approx \frac{2}{\eta R_{\rm cr}^2}.  
 \label{kresult}
\end{equation}
Radius of the critical bubble, $R_{\rm cr}$, depends on one hand on the
cooling rate of the system---in cosmology on the strength of gravitational
interaction---and on the other hand on the thermodynamical properties of the
phase transition, especially on the values of latent heat and surface tension.

\setlength{\unitlength}{0.7mm}
\begin{figure}[t]

\vspace*{-17mm}

\epsfig{file=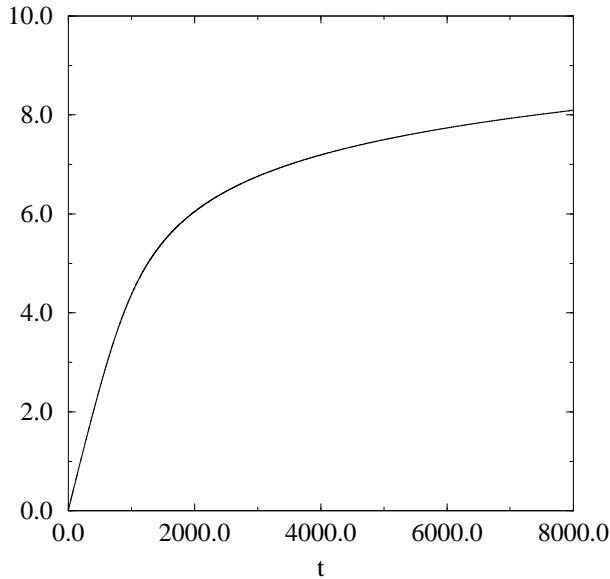,bbllx=20pt,bblly=275pt,bburx=612pt,bbury=792pt,%
 width=11cm,angle=0}
\caption{Numerical determination of the initial growth rate $\kappa$.
Horizontal axis is time in units of inverse thermodynamical phase transition
temperature, $T_c^{-1}$, and vertical axis is
$\ln \{ [R(t)-R_{\rm cr} ] / [R(0)-R_{\rm cr} ]  \} \equiv y$. 
Initial growth rate is given by
$\kappa = \lim_{t \rightarrow 0} \frac{{\rm d}y}{{\rm d}t}$ (when 
$R(0) \rightarrow R_{\rm cr}$). In the figure $\eta=1 T_c$, 
$\xi_{\rm cr}=1.07243 T_c^{-1}$, $R_{\rm cr}=19.695 T_c^{-1}$, 
and $R(0)/R_{\rm cr}=1.01$.
}
\label{fig:kt}
\end{figure}

A more straightforward method is to integrate the dissipation
equation~(\ref{dphidt}) directly in the thin-wall limit.\cite{Kajantie92} The
result for the initial acceleration of the bubble radius is
\begin{equation}
  \frac{{\rm d}R}{{\rm d}t} \approx \frac{2}{\eta R_{\rm cr}^2} \left\{
   \frac{d-1}{2}R - (d-2)R_{\rm cr} \right\}, \label{kd}
\end{equation}
where the dimensionality of space, $d$, is explicitly visible. In the case
$d=3$ Eq.~(\ref{kresult}) follows from this by comparing with
Eq.~(\ref{rkdef}).  Eq.~(\ref{kd}) states clearly how in the real world the
initial growth of bubbles is qualitatively different, much slower, than in one
spatial dimension.

The opposite approach is to let the bubble to expand in a hydrodynamical
computer simulation, and to measure the initial growth rate numerically.
Bubble radius is defined to be the distance where the tension or gradient
energy has the maximum. The value of $\kappa$ can be read from
Fig.~\ref{fig:kt} as the slope of the curve at origin. In the example case
this gives $\kappa = (0.0052 \pm 0.0001)T_c$, which coincides well with the
analytical estimate from Eq.~(\ref{kresult}), $\kappa \approx 0.00516 T_c$.

These results can be applied in the analysis of thermal cosmological phase
transitions. In the case of relativistic heavy-ion collisions there is severe
doubt on the validity of the general framework, nucleation in thermal
systems.

\section*{Acknowledgments}
I would like to thank H. Kurki-Suonio and M. Laine for development of the
spherically symmetric hydrodynamical computer code and for collaboration, and
L. McLerran for discussions.

\end{document}